\documentclass{aa}
\usepackage{graphicx}
\usepackage{psfig}
\begin{document}

\title{UV (IUE) spectra of the  central stars of high latitude planetary
nebulae Hb7 and Sp3
\thanks{ Based on the observations obtained with the International 
Ultraviolet Explorer (IUE), retrieved from the IUE Final Archive at VILSPA,
Madrid, Spain}}
\author{Geetanjali Gauba \inst{1}
\and M. Parthasarathy \inst{1,2}
\and Y.Nakada \inst{3,4}
\and T.Fujii \inst{3}}
\institute{Indian Institute of Astrophysics, Koramangla, Bangalore - 560034, 
India
\and National Astronomical Observatory, 2-21-1 Osawa, Mitaka, Tokyo
181-8588, Japan
\and Institute of Astronomy, School of Science, University of Tokyo, Bunkyo,
Tokyo 113-0033, Japan
\and Kiso Observatory, School of Science, University of Tokyo, Mitaka, Kiso,
Nagano 397-0101, Japan}
\date{Received / accepted}
\authorrunning{Geetanjali Gauba}
\titlerunning{Central stars of planetary nebulae Hb7 and Sp3}

\abstract{
We present an analysis of the UV (IUE) spectra of the central stars of 
Hb7 and Sp3.  Comparison with the IUE spectrum of the standard star 
HD 93205 leads to a spectral classification of O3V for these stars,
with an effective temperature of 50,\,000 K.  From the P-Cygni profiles 
of CIV (1550\AA~), we derive stellar wind velocities and mass loss rates of
$-$1317 kms$^{-1} \pm$ 300 kms$^{-1}$ and 2.9X10$^{-8}$M$_{\odot}$ yr$^{-1}$
and $-$1603 kms$^{-1} \pm $400 kms$^{-1}$ and 7X10$^{-9}$M$_\odot$ yr$^{-1}$ 
for Hb7 and Sp3 respectively. From all the available data, we reconstruct
the spectral energy distribution of Hb7 and Sp3.
}

\maketitle

\section{Introduction}
 
The central stars of planetary nebulae (CSPNe) are in general very hot
objects and their continuum flux is more easily detectable in the UV than 
in the optical. The detection of fast winds in CSPNe from UV observations
may be considered as one of the important discoveries of the IUE satellite
(Heap 1986, Patriarchi and Perinotto 1991).  NV (1240 \AA~) and 
CIV (1550 \AA~) resonance line doublets are the most dominant lines
formed in the winds of hot stars. A study of the stellar wind profiles of
these lines is important to determine the terminal wind
velocities and hence the post-AGB mass-loss rate. 

We have carried out a program to study the wind profiles of several high 
galactic latitude planetary nebulae. A  monitoring of the NV and CIV wind 
profiles in Hen 1357 (=SAO 244567) , showed wind variability in this young 
PN (Parthasarathy et al. 1993, 1995) which may be a signature of 
episodic mass loss in post-AGB stars. In this paper we present an analysis 
of the UV (IUE) low resolution spectra  of the high galactic 
latitude PNe  Hb7 (PN G003.9$-$14.9 = IRAS 18523$-$3219 ; l = 3.97,\, 
b = $-$14.9) and Sp3 (PN G342.5$-$14.3 = IRAS 18033$-$5101; l = 342.51,\,
b = $-$14.32). The photometric colours and optical spectra of these two PNe had 
indicated that they contain hot central stars (Acker et al. 1992, Aller 1976).
We also present the JHK photometry of Hb7 from the 2MASS Point Source Catalog.

\section{Observations}

Low resolution UV spectra of Hb7 and Sp3 were obtained on September 29,
1994 with the SWP camera onboard
the IUE satellite. The SWP52257LL image of Hb7 (80 min exposure) and
the SWP52256LL image of Sp3 (30 min exposure) were obtained by
centering the central stars in the 10\arcsec X 23\arcsec aperture.  
The spectra have been re-extracted from the IUE Final Archive at
VILSPA which were re-processed using the IUE NEWSIPS pipeline 
which applies the SWET extraction method as well as the latest
flux calibration and close-out camera sensitivity corrections.
Line-by-line images have been inspected for spurious features.
  
\section{Analysis} 
The spectra of Hb7 and Sp3 from 1150\AA~ to 1950\AA~ 
in absolute flux units are shown in Fig. 1 a and b. 
The spectra were dereddened by using E(B$-$V) = 0.19 
for Hb7 (from c(H$\beta$) = 0.28, Tylenda et al. 1989), 
and E(B$-$V) = 0.159 for Sp3 (from HST data, Ciardullo et al.\,
1999). The dereddened spectra of Hb7 and Sp3 were compared with 
the dereddened IUE spectra of standard stars (Heck et al. 1984). 
For comparison, the three spectra were normalised at 
$\lambda$ 1601.53\AA~. Savage et al.(1985) quote a value of 
E(B$-$V) = 0.37 for the standard O3 V star, HD93205. Using the
2200\AA~ feature in the UV, we found E(B$-$V) = 0.34 for
this star. Hb7 and Sp3 show good agreement with the UV continuum 
and spectral features of HD93205 (see Fig. 1). Therefore, 
we adopt the same effective temperature of 50,\,000 K for the 
nuclei of Hb7 and Sp3.      

\begin{figure}[h] 
\resizebox{\hsize}{!}{\includegraphics{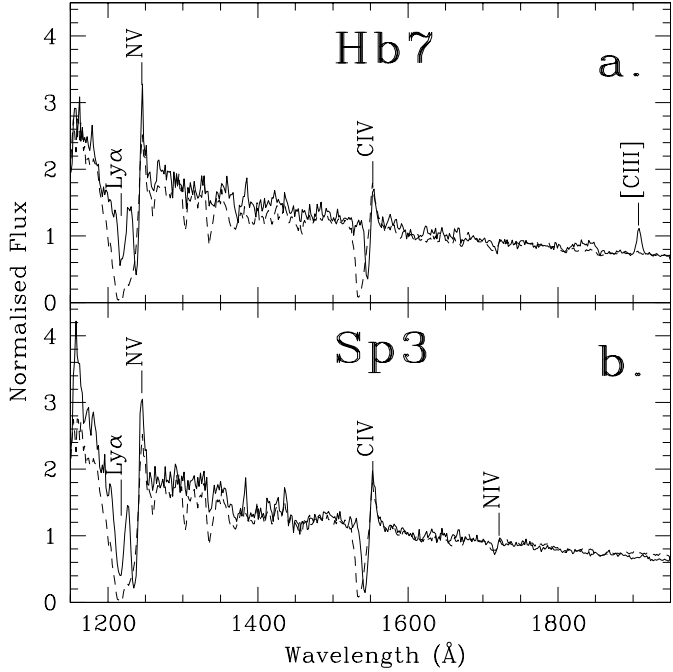}}
\caption{The dereddened spectra of Hb7 {\bf (a)}, E(B$-$V)=0.19 and 
Sp3 {\bf (b)}, E(B$-$V)=0.159 are plotted alongwith the dereddened 
SWP spectra of a standard O3 -dwarf star(HD93205, E(B$-$V)=0.34) from the 
standard star atlas by Heck et al.(1984).  The standard star spectrum 
is represented by a  dotted line. 
}
\end{figure}

\subsection{Terminal Wind Velocity} Using the violet absorption edge in the high resolution UV spectra of HD 93205, Prinja et al. (1990) calculated a terminal wind velocity (v$_{\infty}$) of $-3370$ kms$^{-1}$ for this star. They estimated a measurement error of less than 100 kms$^{-1}$ in the determination of  this Doppler velocity. 

The determination of terminal wind velocity from Doppler
shifts in low-dispersion spectra is complicated by the fact 
that the absorption troughs of  strong (saturated) stellar wind lines 
do not exhibit extended regions of zero residual intensity. 
In the low resolution spectra of Hb7 and Sp3, the violet edge of NV is 
contaminated by Lyman $\alpha$ absorption. By measuring differential
shifts of the CIV absorption profiles of the two stars with respect 
to HD 93205, we found Doppler velocities of $-$1435 kms$^{-1}$ for Hb7
and $-$1628 kms$^{-1}$ for Sp3. These values
may be compared with velocities calculated following the analysis
of Prinja(1994).
 
The empirical relation provided by Prinja (1994) uses the 
difference between the position of the emission peak and the absorption 
minimum for the CIV line i.e.  
v$_{\infty}$ = a$_{1}$ + a$_{2}$($\Delta \lambda$) + a$_{3}$($\Delta \lambda)^
{2}$
where, a$_{1}= -883 \pm$ 48, a$_{2}=  259 \pm$ 9, a$_{3}= -3 \pm$ 2 and
$\Delta \lambda= \lambda^{\rm Emis}_{\rm peak}-\lambda^{\rm Abs}_{\rm min}$,
($\lambda$ in \AA~ and v$_{\infty}$ in kms$^{-1}$). Using this
relation, we 
found v$_{\infty}$= $-$1317 kms$^{-1} \pm$ 316 kms$^{-1}$ for Hb7 and
v$_{\infty}$ = $-$1603 kms$^{-1}\pm$ 389 kms$^{-1}$ for Sp3.
Finally, we adopt a  terminal velocity of 
$-$1317 kms$^{-1} \pm$ 300 kms$^{-1}$ for Hb7 and of
$-$1603 kms$^{-1}\pm$ 400 kms$^{-1}$ for Sp3.

\subsection{Stellar temperature and core-mass}

Samland et al. (1992) had estimated the temperature of the central star of
Hb7 from photoionization model to be 56,\,000K. For the central star of Sp3,
Preite-Martinez et al. (1991) had estimated the energy-balance temperature
to be 39,\,400K.  For an O3 star, the effective temperature (T$_{eff}$) is
estimated to be 50,\,000K (Lang 1992).  We adopt the same value of
50,\,000 K (log Teff = 4.7) for the two stars.\,
Pauldrach et al. (1988) analysed the relation between the effective 
temperature, mass of the nuclei, the terminal velocity and mass-loss 
rate (see their fig. 10 and 6a).  From these relations, we can deduce a 
core-mass of 0.644 and 0.565 M$_\odot$ and a mass-loss rate of 
2.9 $\times$ 10$^{-8}$ M$_{\odot}$ yr.$^{-1}$ and 
7 $\times 10^{-9}$ M$_{\odot}$yr.$^{-1}$ ~for Hb7 
and Sp3 respectively.

\setcounter{table}{0}
\begin{table*} [ht]
\centering
\renewcommand{\thetable}{\arabic{table}}
\caption{Photometric data of Hb7 and Sp3}
\begin{tabular}{lllllll} \hline 

&B&V&I&J&H&K \\ \hline \hline
Hb7&13.76($\Delta$m$<$0.10)&13.97($\Delta$m$<$0.10)&-&12.768$\pm$0.040&12.866
$\pm$0.047&12.251$\pm$0.034\\
& (Tylenda et al. 1989)& (Tylenda et al. 1989)& & ( 2MASS & Point Source & 
Catalog ) \\ \hline 
Sp3&12.45(0.10$<\Delta$m$<$0.25)&13.2($\Delta$m$<$0.05)&13.39($\Delta$m$<$0.05)& - &- &-\\ 
& (Tylenda et al. 1991) & ( Ciardullo  et  al. &1999 ) & & & \\ \hline \hline

\end{tabular}
\end{table*}
\subsection{Spectral Energy Distribution(SED)}

The IUE spectra of Hb7 and Sp3 were combined with the available 
BVI photometry, JHK photometry (Hb7) from the 2MASS Point Source 
Catalog and IRAS photometry at 12, 25, 60 and 100 $\mu$m to reconstruct 
the overall spectral energy distribution (Fig.2 a and b). 
The UV data of both stars shows good agreement with a blackbody
distribution at 50,\,000K. The IRAS fluxes do not seem to obey a single
black body temperature. By fitting  mean blackbody curves to the
IRAS fluxes for Hb7 and Sp3, we estimated cold dust temperatures of
130K and 100K respectively.

The JHK flux distribution, for Hb7, shows no indication of the presence of 
warm dust around the central star.  Warm dust is generally attributed to 
emission from dust grains formed in the outflow close to the central 
star as a result of on-going post-AGB mass loss.  The absence of warm dust 
may be attributed to photodissociation and diffusion of the dust grains formed
close to the hot central star. 

Ciardullo et al. (1999) imaged Sp3 with the Wide Field Planetary Camera 2
onboard HST. They found the central star to be a binary with a separation 
of 0.3\arcsec.  They found V = 13.20, V$-$I = $-$0.19, and E(B$-$V) = 0.159 
for the central star and V=16.86 and V$-$I = 0.83 for the companion. 
They considered it as a probable
physical pair. The binary nature of the nuclei of Sp3, may explain the
too bright value of the magnitude calculated by Tylenda et al.(1991;\,
B = 12.45, V = 12.51 ). The V$-$I colour of the companion is similar to that 
of an F star. The B magnitude by Tylenda et al. (1991) has been corrected
for the contribution from a main sequence F type star and plotted in
Fig. 2b. In the IUE SWP spectrum of Sp3 we do not find any evidence for
the companion star spectrum. Since the F-type companion is several
magnitudes fainter, its effect on the continuum flux
of the central star in the SWP spectrum appears to be insignificant.

Assuming a temperature of 50,\,000K for the central stars, 
the integrated flux in the UV (1150\AA~ to 1950\AA~) is 
1.34$\times10^{-9}$ erg s$^{-1}$ cm$^{-2}$ for Hb7 and 
2.49$\times10^{-9}$erg s$^{-1}$ cm$^{-2}$ for Sp3. The integrated 
far infrared fluxes (12$\mu$ to 100$\mu$) for Hb7 and Sp3 with 
blackbody temperatures of 130K and 100K respectively are 
0.59$\times10^{-9}$ erg s$^{-1}$ cm$^{-2}$
0.45$\times10^{-9}$erg s$^{-1}$ cm$^{-2}$. Thus, almost
as much energy is radiated in the infrared as is seen coming from the
central star(s).  

\noindent In Table 1 we have listed the BVI, JHK magnitudes of Hb7 and Sp3 
adopted in this paper.
The JHK magnitudes for Hb7 were obtained from the
2MASS Point Source Catalog within a search radius of 6 \arcsec.  
The K band image of Hb7 from the 2MASS Catalog is shown in Fig.3.
The NICMOS arrays mounted on the 2MASS telescopes provide a resolution of
2\arcsec per pixel.  Hb7 is not resolved at this resolution and does
not appear as an extended source in the 2MASS JHK images. 

\begin{figure} 
\resizebox{\hsize}{!}{\includegraphics{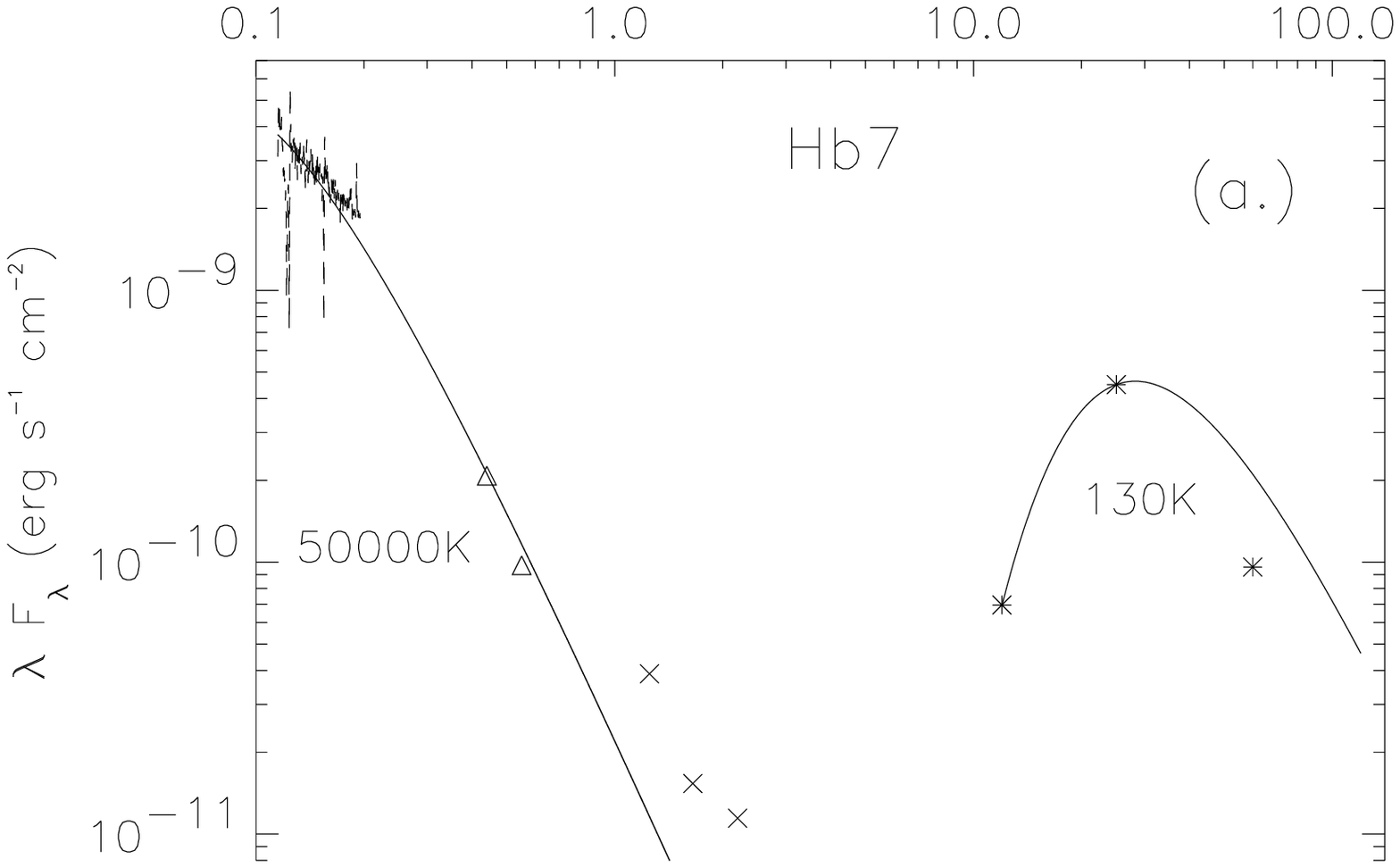}}
\resizebox{\hsize}{!}{\includegraphics{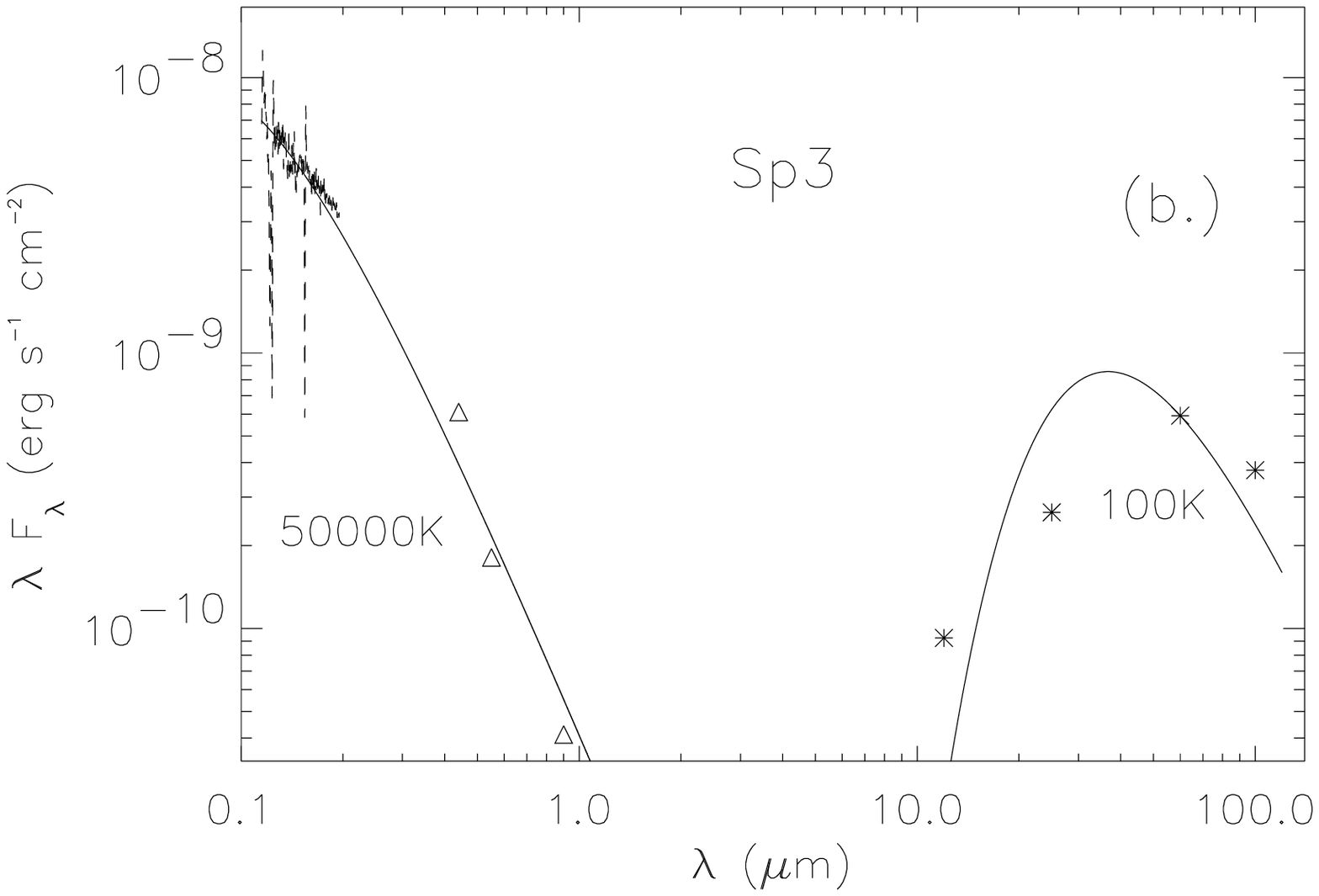}}
\caption{Energy distribution of Hb7 (a) and Sp3 (b) from the UV to the far 
infrared is shown. The data is corrected for interstellar reddening 
using E(B$-$V)=0.19 for Hb7 and E(B$-$V)=0.159 for Sp3. IUE data 
(dashed line) is plotted alongwith BVI (open triangles), JHK (crosses) 
and IRAS photometry (asterix marks).}
\end{figure}

\begin{figure}
\setcounter{figure}{2}
\resizebox{\hsize}{!}{\includegraphics{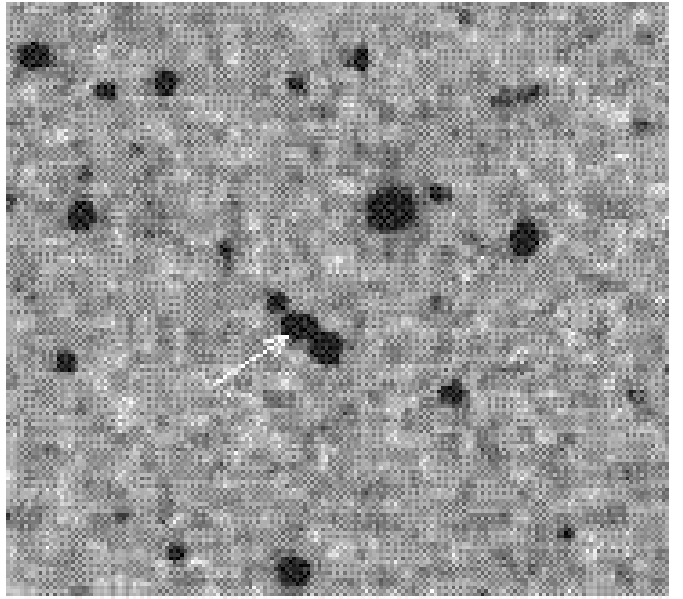}}
\caption{2MASS image of Hb7 in K band}
\end{figure}

\subsection{Dynamical age of the nebula}

For Hb7, all distance estimates in literature have been obtained 
assuming a nebular diameter of
4\arcsec ~(Vorontsov-Velyaminov 1962). However, Vorontsov-Velyaminov's
estimate of the nebular diameter was based on low resolution
photographic plates and may be wrong. Recently based on CCD images, Gorny et
al. (1999) estimated an angular size of 13\arcsec $\times$ 12\arcsec 
~in H$_{\alpha}$. We found this to be consistent with the nebular diameter 
estimated from the 2MASS images. Since distance estimates for 
Hb7 based on the wrong angular diameter of 4\arcsec cannot be
used, we have used the relation
between core-mass and quiescent luminosity
maximum (L$_{\rm Q}$) for AGB stars ( Wood and Zarro 1981) to derive  
the distance. Using L$_{\rm Q}$ for the luminosity 
of the star (8828L$_{\odot}$ for Hb7) and M$_{\rm bol}$(Sun)( = 4.75) 
we found   
M$_{\rm bol}$(Hb7) to be $-$5.1. Applying the bolometric correction and using 
the formula for the distance modulus we obtained a distance of 5.5 kpc.

Gussie and Taylor (1994)  found two components in the expansion velocity
distribution of a large sample of PNe. Nebulae with the low-velocity
component (12.5 kms$^{-1}$) were found to be smaller in linear extent
than high-expansion velocity nebulae (27.5 kms$^{-1}$). 
Assuming an expansion velocity of 12.5 kms$^{-1}$,
angular diameter of 13\arcsec ~and distance of 5.5 kpc, we obtained
a dynamical age of 13418 years for Hb7. 

Using Daub's (1982) formalism and an angular radius of 17.8\arcsec \,
(Acker et al 1992), Cahn et al. (1992) obtained a 
distance of 1.9 $ \pm $ 0.3 kpc for Sp3. The angular diameter and
expansion velocity of Sp3 is 35.5\arcsec ~and 22 kms$^{-1}$
respectively (Acker et al. 1992). At a distance of 1.9 kpc,~
we found the age of the nebula to be 7278 years. 

The theoretical evolutionary tracks of Bl\"ocker and
Sch\"onberner (1990), predict an age of 3000 years
for PNe with log(T$_{eff}$/K) of 4.7 and core mass of 0.605M$_{\odot}$,
along the horizontal part of the evolutionary track on the HR-diagram.

\section{ Discussion and Conclusions}

Our analysis of the UV (IUE) spectra reveals that
the central stars of Hb7 and Sp3 are O3 -dwarfs 
with effective temperatures of 50,\,000K,
core-mass of 0.644 M$_{\odot}$ and 0.565 M$_{\odot}$ and
mass loss rates of 2.9 $\times$ 10$^{-8}$ M$_{\odot}$ yr.$^{-1}$ and
7 $\times 10^{-9}$ M$_{\odot}$ yr.$^{-1}$ respectively. 
The IRAS fluxes of these objects revealed a cold dust component 
at 130K for Hb7 and 100K for Sp3.  The cold dust component may
be interpreted as thermal emission from the dust present in the
circumstellar envelope of these stars, a remnant of the previous strong
mass loss AGB phase. We estimated dynamical ages of 13.4 X 10$^{3}$ years 
and 7.3 X 10$^{3}$ years for Hb7 and Sp3 respectively. 

Cerruti-Sola and Perinotto (1985) investigated the frequency
of occurence of stellar winds in CSPNe. They found that it depends
on the stellar gravity in the sense that CSPNe with a gravity smaller
than log g = 5.2 (cgs) almost always have a wind while at higher gravities
the presence of wind becomes less and less frequent. The presence of wind
in the CSPNe Hb7 and Sp3 indicates that their surface gravities
log g  $<$ 5.2. Pauldrach et al. (1988) have shown that the presence
of a fast wind in a CSPN depends not only on the stellar gravity
but also on the luminosity. That is the more a CSPN departs from the
Eddington luminosity, the less frequent is the occurence of the wind.

\acknowledgements

MP is very thankful to Prof.S.Deguchi, Prof.K.Kodaira and 
Prof. N.Kaifu for their kind support and hospitality. We 
thank the referee for helpful comments.

\end{document}